\documentclass[conference]{IEEEtran}
\IEEEoverridecommandlockouts
\usepackage{cite}
\usepackage{amsmath,amssymb,amsfonts}
\usepackage{graphicx,color,overpic,psfrag}
\usepackage{xcolor}
\usepackage{algorithmic}
\usepackage{graphicx}
\usepackage{textcomp}
\usepackage{fancyhdr}
\usepackage{xcolor}
\usepackage{graphicx}
\usepackage{bm}
\usepackage{subfigure}
\usepackage{multirow}
\usepackage{booktabs}
\usepackage{caption}
\usepackage{booktabs}
\usepackage{amsmath}
\usepackage[numbers,sort&compress]{natbib}
\usepackage{algorithm}
\usepackage{algorithmic}
\usepackage{amsfonts}
\usepackage{hyperref}
\usepackage{xcolor}
\usepackage{tabularx}
\usepackage{enumitem}
\hypersetup{
  colorlinks=true,
  linkcolor=blue,  
  citecolor=blue,  
  urlcolor=blue,   
  linkbordercolor={0 0 1},
  pdfborderstyle={/S/U/W 1} 
}
\usepackage{booktabs}
\usepackage{makecell}

\def\BibTeX{{\rm B\kern-.05em{\sc i\kern-.025em b}\kern-.08em
    T\kern-.1667em\lower.7ex\hbox{E}\kern-.125emX}}

\newcommand{\bl}[1]{\color{black}#1}

\begin{document}
\bstctlcite{IEEEexample:BSTcontrol}

\title{Multimodal-NF: A Wireless Dataset for Near-Field Low-Altitude Sensing and Communications

\author{Mengyuan Li, \textit{Graduate Student Member, IEEE}, Qianfan Lu, Jiachen Tian, Hongjun Hu, Yu Han, \textit{Member, IEEE},\\ Xiao Li, \textit{Senior Member, IEEE}, Chao-Kai Wen, \textit{Fellow, IEEE}, and Shi Jin, \textit{Fellow, IEEE}\\
}
\thanks{M. Li, Q. Lu, J. Tian, H. Hu, Y. Han, X. Li, and S. Jin are with the School of Information Science and Engineering, Southeast University, Nanjing 210096, China (email: mengyuan$\_$li@seu.edu.cn; qianfan$\_$lu@seu.edu.cn; huhongjun@seu.edu.cn; tianjiachen@seu.edu.cn; li\_xiao@seu.edu.cn; hanyu@seu.edu.cn; jinshi@seu.edu.cn). 

C.-K. Wen is with the Institute of Communications Engineering, National Sun Yat-sen University, Kaohsiung 804, Taiwan (e-mail: chaokai.wen@mail.nsysu.edu.tw).
}
}
\maketitle

\begin{abstract}
Environment-aware 6G wireless networks demand the deep integration of wireless and multimodal data. However, most existing datasets are confined to 2D terrestrial far-field scenarios, lacking the 3D spatial context and near-field characteristics crucial for low-altitude extremely large-scale multiple-input multiple-output (XL-MIMO) systems. {\bl To bridge this gap, this letter introduces Multimodal-NF, a wireless-centric, large-scale multimodal dataset and the generator framework.} Operating in the upper midband, it synchronizes high-fidelity near-field channel state information (CSI) and precise wireless labels (e.g., Top-5 beam indices and {\bl a binary LoS indicator}) with comprehensive sensory modalities (RGB images, LiDAR point clouds, and GPS). Crucially, these multimodal priors provide spatial semantics that help reduce the near-field search space and thereby lower the overhead of wireless sensing and communication tasks. Finally, we validate the dataset through representative case studies, demonstrating its utility and effectiveness. The open-source generator and dataset are available at \url{https://github.com/Lmyxxn/Multimodal-NF}.
\end{abstract}


\section{Introduction} 

\IEEEPARstart{T}{owards} sixth-generation (6G) wireless networks, extremely large-scale multiple-input multiple-output (XL-MIMO) and the emerging low-altitude economy (LAE) have become key driving forces~\cite{wu2026low}. Building upon the momentum of the upper 6 GHz (U6G) band, extending into the broader upper midband (7-24 GHz) serves as a strategic spectrum choice for these 3D low-altitude systems~\cite{Zhang2025NewMidband}. Its short wavelengths allow for dense XL-MIMO packing, while contiguous wide bandwidths significantly boost transmission rates~\cite{Tian2025MidBand}. Although artificial intelligence (AI) has demonstrated immense potential in optimizing communication systems~\cite{Zhang2026ComAI}, its efficacy fundamentally relies on massive training data. The lack of large-scale, high-quality datasets that capture the complex 3D near-field characteristics of upper midband LAE systems creates a severe bottleneck.

Existing representative wireless datasets and generators (summarized in Table \ref{tab:dataset_comparison}) face critical limitations for LAE applications. Real-world datasets (e.g., DeepSense6G \cite{alkhateeb2023deepsense}) suffer from rigid physical settings, while prevailing simulators (e.g., DeepMIMO \cite{alkhateeb2019deepmimo}) rely on far-field planar wave assumptions or lack 3D dynamic multimodal support. Even recent near-field models \cite{yu2025buptcmcc} remain confined to 2D terrestrial environments.
To bridge this gap, we introduce Multimodal-NF, with the main contributions summarized as follows:

\begin{table*}[t!]
\centering
\caption{Comparison of Representative Wireless Datasets and Generation Frameworks.}
\label{tab:dataset_comparison}
\resizebox{\textwidth}{!}{%
\renewcommand{\arraystretch}{0.9} 
\setlength{\tabcolsep}{3pt} 
\scriptsize 
\begin{tabular}{@{}llcccccc@{}} 
\toprule
\textbf{Category} & \textbf{Dataset/Framework} & \textbf{BS Antenna} & \textbf{Frequency Band} & \textbf{Near-Field} & \textbf{3D} & \textbf{Customizable} & \textbf{Modalities} \\ 
\midrule

\multirow{2}{*}{\shortstack{Real-World\\Measured}} 
& DeepSense 6G \cite{alkhateeb2023deepsense}  & 16 ULA             & mmWave & $\times$ & $\checkmark$ & $\times$ & CSI, LiDAR, Radar, RGB, GPS \\
& LuViRA \cite{yaman2024luvira}               & $25 \times 4$ UPA  & Mid-band & $\times$ & $\times$ & $\times$ & CSI, RGB, Depth, IMU, Audio \\ 

\midrule

\multirow{6}{*}{\shortstack{Simulated}} 
& DeepMIMO \cite{alkhateeb2019deepmimo}       & $16 \times 16$ UPA & mmWave & $\times$ & $\times$ & $\times$ & CSI only \\
& BUPTCMCC-6G \cite{yu2025buptcmcc}           & Custom             & Upper midband, mmWave, THz & $\checkmark$ & $\times$ & $\checkmark$ & CSI only \\
& Multimodal Wireless \cite{mao2025multimodal}& UPA ($\le 8 \times 8$) & mmWave & $\times$ & $\times$ & $\checkmark$ & CSI, LiDAR, Radar, RGB, Depth, IMU \\ 
& Raymobtime \cite{klautau20185g}             & $4 \times 4$ UPA   & mmWave & $\times$ & $\times$ & $\times$ & CSI, LiDAR, Radar, RGB \\
& \textbf{Multimodal-NF}                      & \textbf{$64 \times 64$ UPA} & \textbf{U6G, upper midband} & $\checkmark$ & $\checkmark$ & $\checkmark$ & \textbf{CSI, RGB, LiDAR, GPS, Wireless Labels} \\

\bottomrule
\end{tabular}%
}
\end{table*}

\begin{itemize}[leftmargin=*, nosep]
    \item \textbf{Customizable 3D Near-Field Generator:} We develop an open-source dataset generator for low-altitude XL-MIMO systems, which allows researchers to flexibly define 3D UAV trajectories, array configurations, and environments.
    
    \item \textbf{{\bl Wireless-Centric} Multimodal Dataset:} We provide a large-scale dataset that synchronizes high-fidelity wireless data (e.g., near-field channel state information (CSI), Top-5 beam indices, line-of-sight (LoS)/non-LoS (NLoS) labels) with rich sensory information, including RGB images, LiDAR point clouds, and GPS coordinates.
    
    \item \textbf{Dataset Analysis and Validation:} We analyze and validate the dataset through representative case studies. These results demonstrate its efficacy in empowering environment-aware sensing and communications.
\end{itemize}

\section{System Model}
\label{sec:System Model}

\subsection{Near-Field Channel Model}

\begin{figure}[t]
\centering
\includegraphics[width=0.85\linewidth]{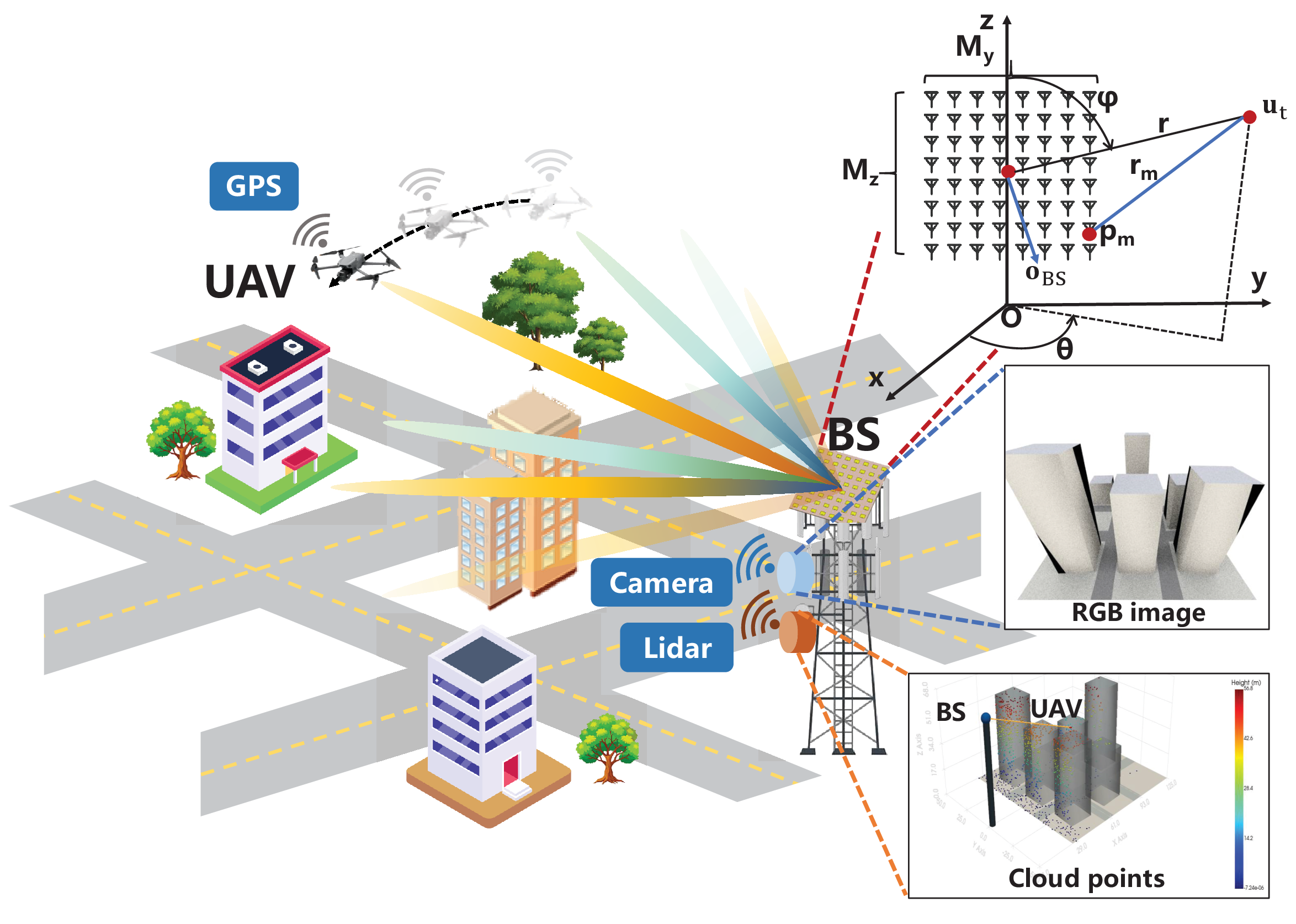}
\caption{{\bl Illustration of the low-altitude XL-MIMO system.}}
\label{fig:LAE_system}
\end{figure}

As depicted in Fig.~\ref{fig:LAE_system}, we investigate a low-altitude XL-MIMO system where a base station (BS) is equipped with an $M$-element ($M = M_y \times M_z$) uniform planar array (UPA) in the $yz$-plane with half-wavelength spacing. A single-antenna unmanned aerial vehicle (UAV), acting as the user equipment (UE), moves in the system. {\bl In the considered coordinate system, $\varphi$ denotes the zenith-angle measured from the $+z$-axis, and $\theta$ denotes the azimuth defined in the $xy$-plane, where $[0^\circ, 90^\circ]$ spans from the $+x$-axis to the $+y$-axis, and $[-90^\circ, 0^\circ]$ spans from the $-y$-axis to the $+x$-axis.}
To ensure cross-modal spatial alignment, an RGB camera and a LiDAR sensor are co-located with the UPA at a coordinate of $\mathbf o_{\mathrm {BS}}=[0,0,h]^{\mathsf T}$, oriented towards the $+x$-axis, sharing the same field-of-view (FoV) and global coordinate system. The UAV's ground-truth (GT) 3D spatial coordinate at time $t$ is denoted by $\mathbf{u}_t \in \mathbb{R}^3$.
The near-field uplink channel between the $m$-th BS antenna at $\mathbf{p}_{m}$ and the UE at time $t$ and frequency $f$ can be expressesd as
\begin{equation}
    \label{eq:channel_model}
    h_{m}(t, f) = \sum_{l=1}^{L(t)} g_{l,m}(t, f) e^{-j \frac{2\pi f}{c} d_{l,m}(t)},
\end{equation}
where $L(t)$, $g_{l,m}(t,f)$, and $d_{l,m}(t)$ denote the number of paths, the complex gain, and the propagation distance of the $l$-th path ending at the $m$-th antenna, respectively. $c$ is the speed of light. This per-antenna distance computation explicitly captures the near-field spherical wavefront. The uplink received signal can be expressed as

\begin{equation}
    y(t, f) = \mathbf{w}^{\mathsf{H}} \left( \sqrt{P_r}\mathbf{h}(t, f) + \mathbf{n}(t, f) \right),
\end{equation}
where $\mathbf{w}$ is the beamforming vector, $P_r$ is the receive power, and $\mathbf{n}(t, f) \sim \mathcal{CN}(\mathbf{0}, \sigma^2\mathbf{I})$ is the additive Gaussian noise.

\subsection{Motivation for Multimodal Support}

We use $\mathbf{x}_t=(\mathbf{u}_t,\mathcal{E})$ to denote the underlying geometric state (i.e., UAV location $\mathbf{u}_t$ and propagation environment $\mathcal{E}$), which governs both sensing and communication targets, i.e., $\mathbf{s}_t$ and $\mathbf{c}_t$. Recovering these targets from pure wireless data is inherently uncertain due to the high-dimensional 3D low-altitude near-field regions. However, by introducing multimodal observations $V_t$, we can reduce the uncertainty of the geometric state. Following the chain rule of conditional entropy, the residual uncertainty of the communication task satisfies:
$H(\mathbf{c}_t \mid V_t) \le H(\mathbf{x}_t \mid V_t) + H(\mathbf{c}_t \mid \mathbf{x}_t).$
A similar bound holds for the sensing task $\mathbf{s}_t$. This inequality reveals a shared mechanism: informative multimodal data $V_t$ reduces the geometric uncertainty $H(\mathbf{x}_t \mid V_t)$, which in turn bounds the residual uncertainty of both tasks, effectively reducing their search spaces. Since deriving these optimal mappings is analytically intractable in real-world environments, data-driven learning is essential, which fundamentally motivates the construction of the proposed \textit{Multimodal-NF} dataset.

\begin{figure}[t]
\centering
\includegraphics[width=0.9\linewidth]{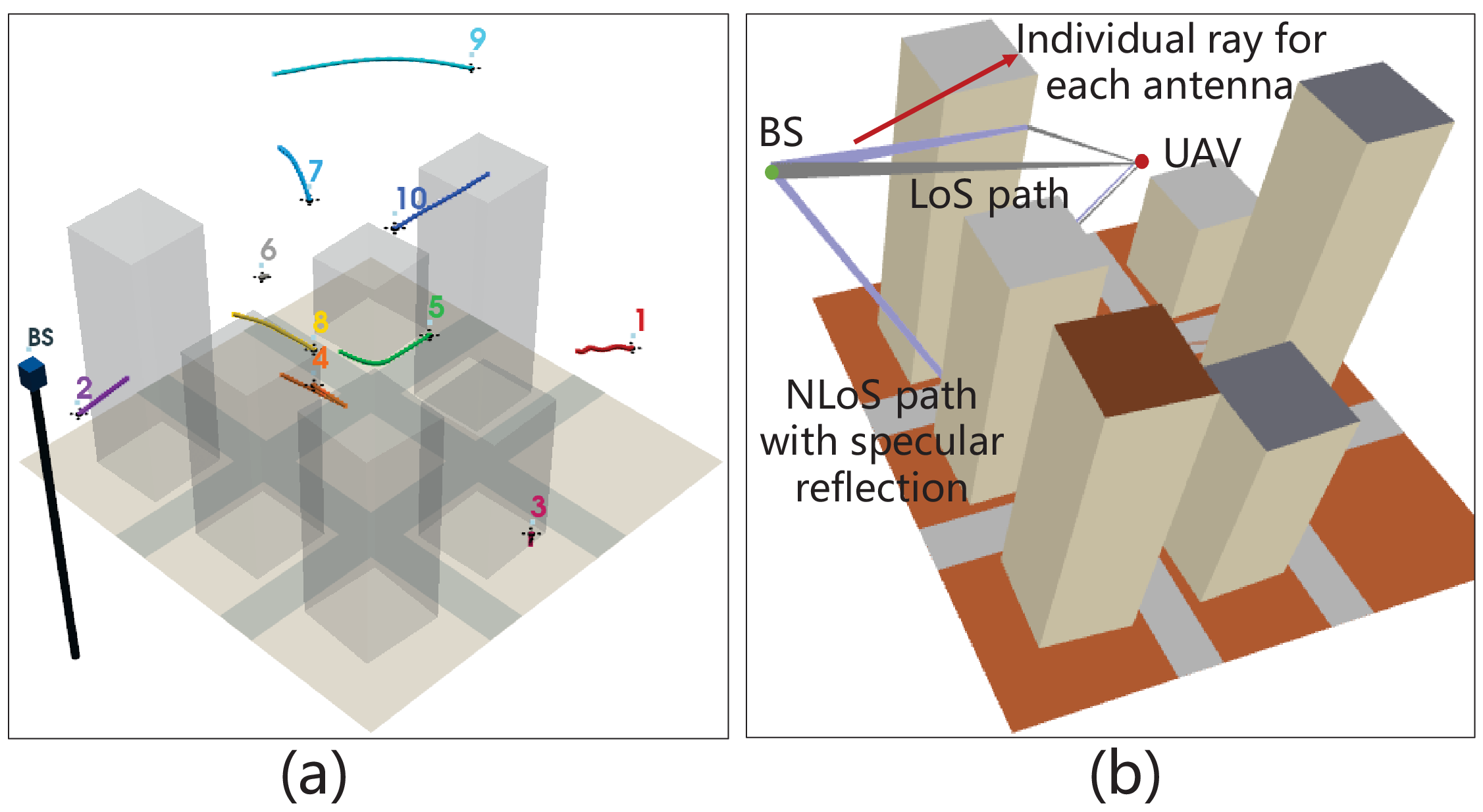}
\caption{{\bl Example visualizations of (a) the LAE scene with the pre-defined trajectory modes and (b) the simulated near-field rays.}}
\label{fig:scene_and_RT}
\end{figure}

\section{Multimodal-NF: Dataset Generator}
\label{sec:Multimodal-NF: Dataset Generator}
In this section, we detail the workflow of the proposed generator, including: (i) scene construction, (ii) trajectory simulation, and (iii) wireless and multimodal data generation.

\subsection{Scene Construction and Trajectory Simulation}
As illustrated in Fig.~\ref{fig:scene_and_RT}(a), the simulation scene spans a $120\,\text{m} \times 120\,\text{m}$ observed region, with the BS deployed at a fixed height on the scene boundary. The scene allows for flexible configuration of road topologies, building densities, and geometric attributes, where building heights are realistically constrained between $20\,\text{m}$ and $60\,\text{m}$. To facilitate channel generation via the Sionna ray-tracing (RT) engine, we assign specific electromagnetic materials to all 3D geometric entities according to ITU-R Recommendation P.2040-3~\cite{itu_p2040_3}. These primarily include ITU-concrete for roads, ITU-marble/wood/metal for buildings, and medium dry ground for the terrain, which provide physically meaningful reflection properties for the Sionna RT process. {\bl In the dataset generation, LoS propagation and specular reflections are enabled with a maximum interaction depth of three, while diffraction and diffuse scattering are not enabled in the current version. This setting is adopted because the considered low-altitude upper-midband channels are mainly dominated by LoS and specular reflection components, whereas diffraction and diffuse scattering require more fine-grained geometry and material calibration and would significantly increase the computational cost of large-scale per-antenna near-field ray tracing.}

The proposed dataset observes $T$ consecutive time slots for each UAV. To reflect typical low-altitude operations, we incorporate 10 trajectory modes representing diverse kinematic behaviors. As detailed in Table~\ref{tab:trajectory_modes}, each mode is characterized by horizontal and vertical velocity ranges as well as the trajectory descriptions.
Furthermore, these trajectories are categorized into two difficulty levels: hard and easy. The hard modes (\textit{Zigzag}, \textit{Sudden Turn}, \textit{Wall Hug}, and \textit{Inspect}) introduce challenging propagation conditions. Specifically, highly mobile modes, such as \textit{Zigzag} and \textit{Sudden Turn}, are designed to test the ability to track rapid spatial variations, while \textit{Wall Hug} and \textit{Inspect} simulate NLoS-dominant environments to evaluate robustness against blockages. Conversely, the easy modes (\textit{Street Patrol}, \textit{City Cruise}, \textit{Orbit}, \textit{Scan}, \textit{Fast Transit}, and \textit{Hover}) feature relatively stable flights.

\definecolor{cZigzag}{HTML}{D50000}
\definecolor{cWallHug}{HTML}{AA00FF}
\definecolor{cInspect}{HTML}{C51162}
\definecolor{cSuddenTurn}{HTML}{FF6D00}
\definecolor{cStreetPatrol}{HTML}{00C853}
\definecolor{cHover}{HTML}{9E9E9E}
\definecolor{cCityCruise}{HTML}{00B0FF}
\definecolor{cOrbit}{HTML}{FFD600}
\definecolor{cFastTransit}{HTML}{00E5FF}
\definecolor{cScan}{HTML}{2962FF}

\begin{table}[t]
    \centering
    \caption{{\bl Characteristics of defined UAV trajectory modes.}}
    \label{tab:trajectory_modes}
    
    \footnotesize 
    
    \renewcommand{\arraystretch}{1.2} 
    \setlength{\tabcolsep}{2pt} 

    \begin{tabular}{@{}cl c cl@{}}
        \toprule
        \textbf{ID} & \textbf{Trajectory} & 
        \makecell[c]{\textbf{Vel. (m/s)} \\ \textbf{Horiz. / Vert.}} & 
        \makecell[c]{\textbf{Altitude} \\ \textbf{(m)}} & 
        \textbf{Description} \\
        \midrule
        1 & \textcolor{cZigzag}{\textbf{Zigzag}}          & 0--5 / 0--1.5   & 5--15  & Periodic lateral weaving motion \\
        2 & \textcolor{cWallHug}{\textbf{Wall Hug}}       & 5--15 / 0       & 5--20  & Building perimeter tracking \\
        3 & \textcolor{cInspect}{\textbf{Inspect}}        & 0 / 0--2        & 2--60  & Vertical facade scanning \\
        4 & \textcolor{cSuddenTurn}{\textbf{Sudden Turn}} & 8--12 / 0--2    & 5--45  & Street flight with abrupt turns \\
        5 & \textcolor{cStreetPatrol}{\textbf{Street Patrol}}& 8--12 / 0--2 & 5--45  & Road-network patrol \\
        6 & \textcolor{cHover}{\textbf{Hover}}            & 0 / 0--0.5      & 10--80 & Quasi-stationary 3D drift \\
        7 & \textcolor{cCityCruise}{\textbf{City Cruise}} & 8--15 / 0       & 30--60 & Smooth linear crossing \\
        8 & \textcolor{cOrbit}{\textbf{Orbit}}            & 0--10 / 0       & 30--60 & Circular flight around building \\
        9 & \textcolor{cFastTransit}{\textbf{Fast Transit}}& 15--25 / 0     & 50--80 & High-speed transit \\
        10 & \textcolor{cScan}{\textbf{Scan}}             & 0--12 / 0       & 50--80 & Back-and-forth grid sweeping \\
        \bottomrule
    \end{tabular}
\end{table}

\subsection{Wireless and Multimodal Data Generation}

\subsubsection{Wireless Data}

We configure the Sionna RT simulator with \texttt{synthetic\_array=False} to enforce per-antenna channel generation. By calculating element-wise propagations, this approach accurately captures the spherical wavefront characteristics inherent to near-field channels, as visualized by individual ray paths in Fig.~\ref{fig:scene_and_RT}(b), and the generation details can be found in~\cite{sionnaRT}. Stacking the channels across all $M$ antennas yields $\mathbf{H}(t)\in\mathbb{C}^{M\times K}$ over $K$ subcarriers, and the CSI over a trajectory of $T$ frames is stored in the released dataset by separating the real and imaginary parts into the CSI tensor $\widetilde{\mathbf{H}} \in \mathbb{R}^{M\times K\times T \times 2}$. {\bl The generated channels focus on the over-the-air near-field propagation process, without modeling array-level electromagnetic or hardware-related effects such as mutual coupling, antenna mismatch, or RF-chain non-idealities. These effects can be incorporated in future work through post-processing array-response models.} {\bl We also provide supplementary wireless labels for downstream tasks: (i) a binary LoS indicator, where a sample is labeled as LoS if its multipath channel contains a direct BS-UE LoS component, and as NLoS otherwise}, (ii) the Top-5 optimal beam indices, and (iii) the corresponding normalized beamforming gains.

{\bl To derive these beam labels, we construct a compact and computationally efficient 3D near-field codebook $\mathcal{W}$ by uniformly sampling the dominant propagation region, rather than the entire three-dimensional space, across azimuth $\theta_{k_\theta} \in [-72^\circ, 72^\circ]$, zenith angle $\varphi_{k_\varphi} \in [60^\circ, 150^\circ]$, and distance $d_{k_r} \in [20, 155]\text{ m}$, using $N_\theta = 20$, $N_\varphi = 20$, and $N_r = 10$ grid points.
Following the coordinate convention in Sionna RT, the sampled polar tuple 
$(\theta_{k_\theta},\varphi_{k_\varphi},d_{k_r})$ is converted into Cartesian coordinate as
\begin{equation}
\mathbf p_{\mathrm{cw}}=\mathbf o_{\mathrm {BS}}+d_{k_r}\big[\sin\varphi_{k_\varphi}\cos\theta_{k_\theta},\,\sin\varphi_{k_\varphi}\sin\theta_{k_\theta},\,\cos\varphi_{k_\varphi}\big]^{\mathsf T},
\end{equation}}
Based on this conversion, the near-field codeword is
\begin{equation}
    \mathbf{w} = \frac{1}{\sqrt{M}} \Big[ e^{-j \frac{2\pi f_c}{c} \|\mathbf{p}_{\text{cw}} - \mathbf{p}_{1}\|}, \dots, e^{-j \frac{2\pi f_c}{c} \|\mathbf{p}_{\text{cw}} - \mathbf{p}_{M}\|} \Big]^{\mathsf{T}},
\end{equation}
where $f_c$ denotes the carrier frequency. For a codeword $\mathbf{w} \in \mathcal{W}$ at time slot $t$, the achievable rate is
\begin{equation} 
    R(\mathbf{w}, t, f_c) = \log_2\!\left( 1 + \frac{P_r \left|\mathbf{w}^{\mathsf{H}}\mathbf{h}(t, f_c)\right|^2}{\sigma^2}\right).
\label{eq:achievable_rate}
\end{equation}
Each beam can be uniquely identified by a global index $k$:
\begin{equation}
    k = (k_\theta - 1) N_\varphi N_r + (k_\varphi - 1) N_r + k_r.
\label{eq:global_beamindex}
\end{equation}
We select the Top-5 global indices that maximize $R(\mathbf{w}, t, f_c)$ and map them back to their spatial tuples $(k_\theta, k_\varphi, k_r)$ to generate the GT Top-5 decomposed labels. To evaluate beam alignment, the normalized beamforming gain is defined as
\begin{equation}
    G_{\text{norm}}(\mathbf{w}, t, f_c)
= \frac{\left|\mathbf{w}^\mathsf{H}\mathbf{h}(t, f_c)\right|^2}
{\left|\mathbf{w}_{\rm opt}^\mathsf{H}\mathbf{h}(t, f_c)\right|^2},
\label{eq:normalized_beam_gain}
\end{equation}
where $\mathbf{w}_{\rm opt}$ denotes the optimal GT beamforming vector.

\begin{table}[t]
\centering
\caption{Contents of Multimodal-NF Dataset.}
\label{tab:dataset_summary}
\small 
\setlength{\tabcolsep}{3pt} 
\renewcommand{\arraystretch}{0.9} 
\scriptsize 
\begin{tabular}{lp{6.8cm}}
\toprule
\textbf{Modality} & \textbf{Data Components \& Attributes} \\
\midrule
Wireless & CSI tensor $\widetilde{\mathbf{H}} \in \mathbb{R}^{M \times K \times T\times2}$ (real/imaginary stacked), {\bl Binary LoS Indicator (1: LoS path exists; 0: otherwise)}, Top-5 Beam Indices, Normalized Beamforming Gains\\
GPS & 3D Coordinates with Gaussian noise $\mathcal{N}(0, \sigma_{\text{GPS}}^2\mathbf{I})$ \\
Vision & RGB Image (FoV=$90^\circ$, $512 \times 512$) \\
LiDAR & 10,000-point Cloud (co-located with camera) \\
Label & Trajectory ID (10 kinematic modes) \\
\bottomrule
\end{tabular}
\end{table}

{\bl
\subsubsection{Multimodal Sensing Data}
To provide multimodal side information for wireless tasks, we generate synchronized RGB images, LiDAR points, and GPS positions for each frame, with detailed configurations and data dimensions summarized in Table~\ref{tab:dataset_summary}. These modalities are frame-aligned with the wireless samples, ensuring that each time slot corresponds to the same UAV state and propagation environment.
Specifically, RGB images are rendered from the BS-side viewpoint, while a co-located LiDAR sensor is simulated using Open3D~\cite{Zhou2018}. The RGB modality captures visual semantics such as scene appearance and object visibility, whereas LiDAR provides 3D structural information of the environment and the UAV. Representative visualizations are shown in Fig.~\ref{fig:LAE_system}. To emulate practical positioning measurements, we corrupt the GT coordinate with Gaussian noise to obtain
$
\tilde{\mathbf{u}}_t = \mathbf{u}_t + \mathbf{z}_t
$,
where $\mathbf{z}_t \sim \mathcal{N}(\mathbf{0}, \sigma_{\text{GPS}}^2 \mathbf{I})$. The resulting GPS observation offers a coarse spatial prior that can be fused with RGB and LiDAR to support downstream wireless tasks. Each sample is also annotated with a trajectory mode ID. More realistic RGB and LiDAR degradations, e.g., due to weather and illumination variations, will be explored in future work.
}

\section{Dataset Analysis and Case Studies}
\label{sec:Dataset Analysis and Case Study}

In this section, we validate the proposed Multimodal-NF dataset through two representative case studies: near-field localization and multimodal beam prediction. To reflect representative upper mid-band operating conditions, the simulation operates at $f_c = 7$ GHz with $\Delta f = 30$ kHz across $K = 128$ subcarriers. The BS employs a $64 \times 64$ vertically polarized UPA. Detailed dataset splits and statistics are summarized in Table~\ref{tab:dataset_stats}. {\bl The average number of paths is $2.53$, and the average root-mean-square (RMS) delay spread is $2.17$~ns.} For downstream evaluation, we utilize single-carrier CSI at $f_c$ and set the GPS noise variance to $\sigma_{\text{GPS}}^2 = 0.5$.

\begin{table}[t]
\centering
\caption{Summary of the Multimodal-NF dataset splits and sample statistics (UPA $64 \times 64$).}
\label{tab:dataset_stats}
\setlength{\tabcolsep}{5pt}
\renewcommand{\arraystretch}{1.0}
\footnotesize
\begin{tabular}{llccccc}
\toprule
\multirow{2}{*}{\textbf{Split}} & \multirow{2}{*}{\textbf{Mode}} & \multirow{2}{*}{\textbf{Cities}} & \multirow{2}{*}{\textbf{Traj.}} & \multicolumn{3}{c}{\textbf{Samples}} \\
\cmidrule(lr){5-7}
 &  &  &  & \textbf{Total} & \textbf{LoS} & \textbf{NLoS} \\
\midrule

\multirow{2}{*}{\textbf{Train}} 
 & Easy & \multirow{2}{*}{22} & 2,614 & 52,280  & 49,923  & 2,357 \\
 & Hard &                     & 5,185 & 103,700 & 94,251  & 9,449 \\
\midrule

\multirow{2}{*}{\textbf{Val}} 
 & Easy & \multirow{2}{*}{4} & 488   & 9,760   & 9,374   & 386 \\
 & Hard &                    & 988   & 19,760  & 19,139  & 621 \\
\midrule

\multirow{2}{*}{\textbf{Test}} 
 & Easy & \multirow{2}{*}{4} & 494   & 9,880   & 9,381   & 499 \\
 & Hard &                    & 1,001 & 20,020  & 19,007  & 1,013 \\
\midrule

\textbf{Total} & -- & \textbf{30} & \textbf{10,770} & \textbf{215,400} & \textbf{201,075} & \textbf{14,325} \\
\bottomrule
\end{tabular}

\vspace{2pt}
\begin{minipage}{0.98\columnwidth}
\footnotesize
\textit{Note:} All trajectories contain $T=20$ frames with a sampling interval of 0.1\,s. The BS UPA is located at $(0,0,65)$\,m. The dataset is split by cities into Train/Val/Test = 22/4/4. Overall, 93.35\% of samples are LoS and 6.65\% are NLoS, {\bl which reflects typical low-altitude communication environments.}
\end{minipage}
\end{table}

\begin{figure}[t]
\centering
\includegraphics[width=0.85\linewidth]{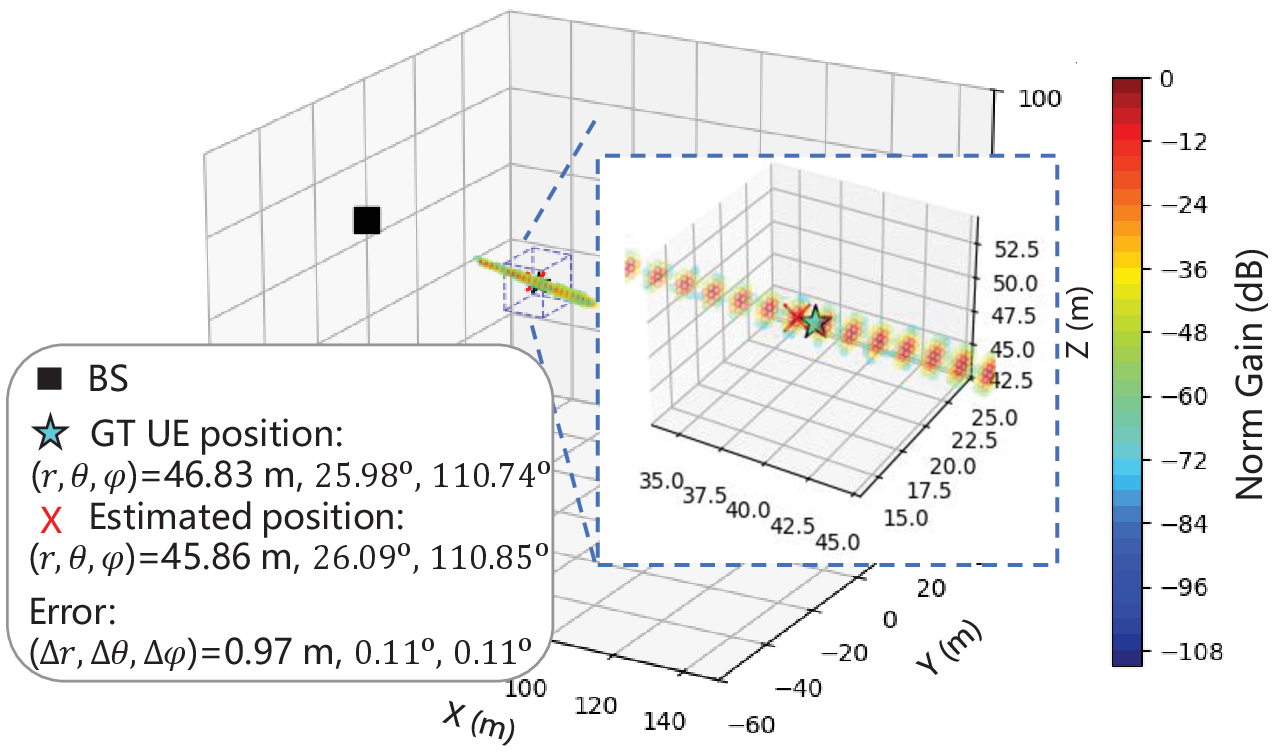}
\caption{{\bl Example of the Cartesian-domain channel energy distribution with the 3D GT UE position and estimated position, where warmer/brighter colors indicate stronger channel energy.}}
\label{fig:loc_est_vis}
\end{figure}

\begin{figure*}[t]
\centering
\includegraphics[width=0.62\linewidth]{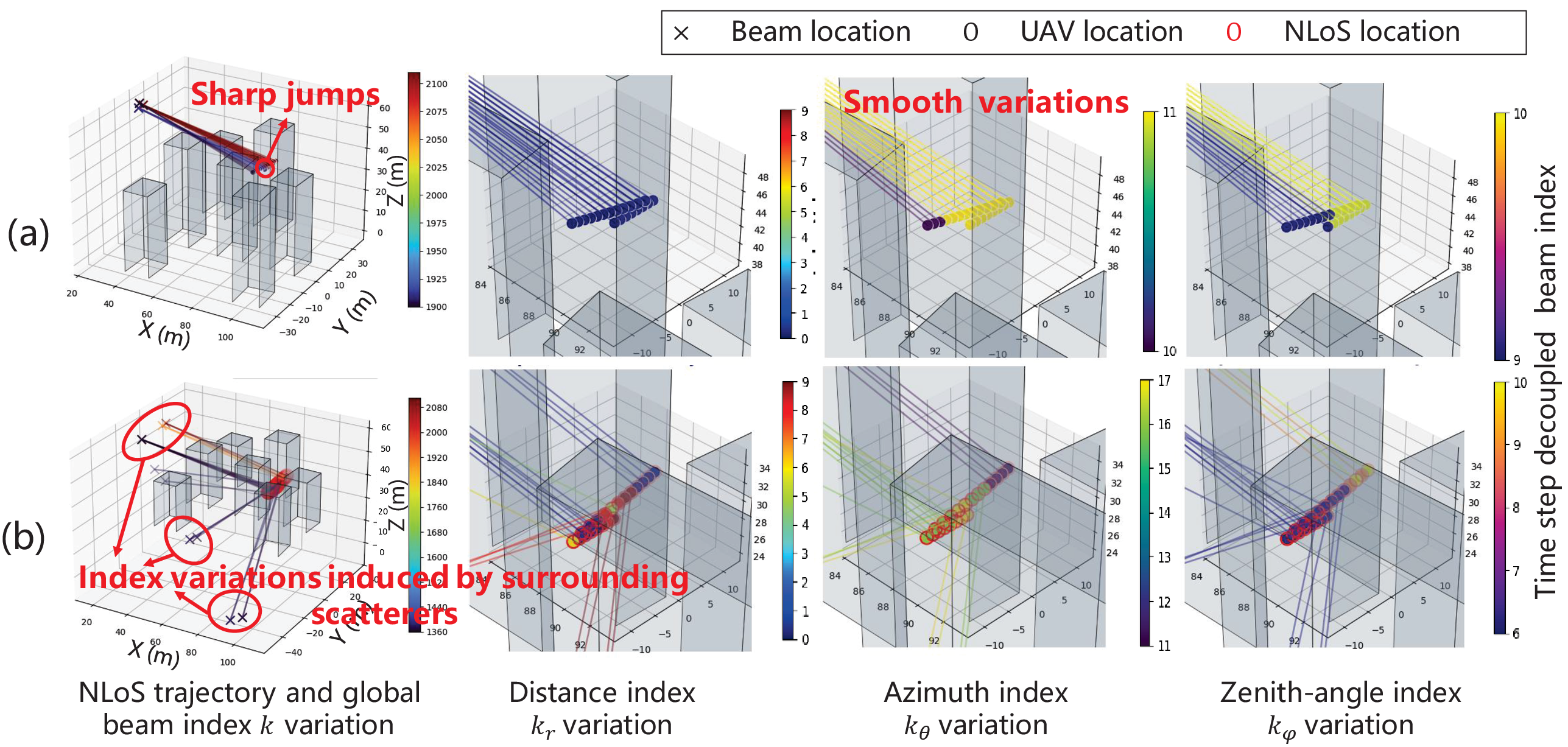}
\caption{{\bl Spatial-temporal variation of beam indices in (a) LoS and (b) NLoS scenarios, including the global index variations and decomposed distance $d$, azimuth $\theta$, and zenith-angle $\varphi$ indices over time. The color indicates the beam index value.}}
\label{fig:beamindex_vis}
\end{figure*}

{\bl 
\subsubsection{Case Study I: User Localization}

To evaluate the proposed dataset, we first conduct 3D near-field user localization using the classic orthogonal matching pursuit (OMP) algorithm~\cite{tropp2007signal} with a fine-grained polar-domain codebook defined in~\cite{Li_KeypointNF_Localization_2025} with $N_\theta = 70$, $N_\varphi = 60$, and $N_d=60$. On the test dataset, the average 3D position error of the OMP-based benchmark is $3.29$~m. The corresponding average distance, azimuth, and zenith-angle errors are $3.01$~m, $0.87^\circ$, and $0.66^\circ$, respectively. As illustrated in Fig.~\ref{fig:loc_est_vis}, the estimated UE positions and angles closely follow the GT, indicating that the generated CSI preserves distinguishable near-field angle--distance characteristics.
We further evaluate a learning-based trajectory prediction baseline proposed in~\cite{Li2026BeamLLM}. By exploiting historical multimodal observations and predicted beam-domain evolution, this method achieves an average 3D position error of $1.92$~m, with average distance, azimuth and zenith-angle errors of $1.84$~m, $0.76^\circ$ and $0.59^\circ$, respectively.
 These results validate the applicability of the proposed dataset to near-field sensing-related research.}

\begin{figure}[t]
\centering
\includegraphics[width=0.6\linewidth]{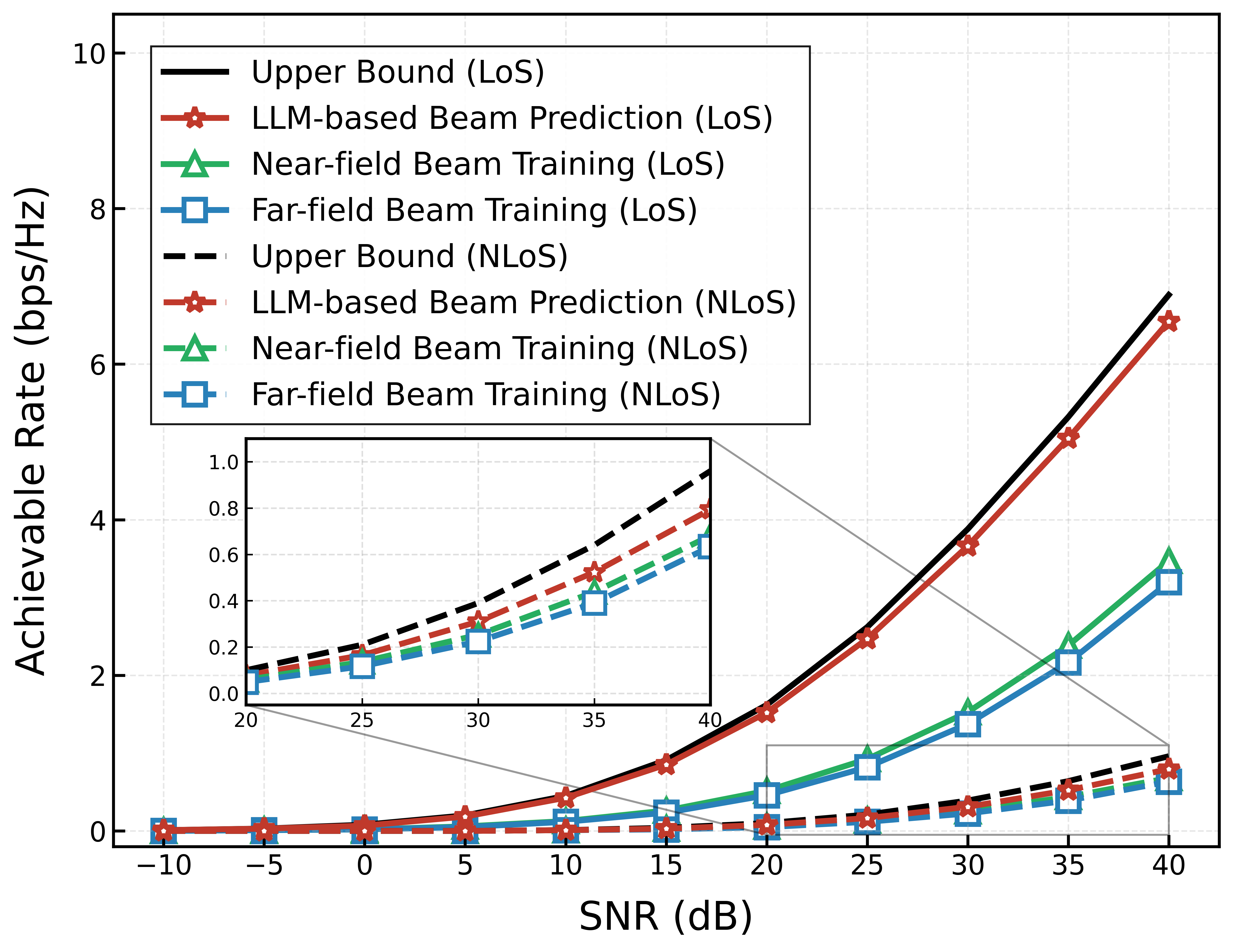}
\caption{System achievable rate comparison of the LLM-based beam prediction method (trained on the proposed dataset) with traditional beam training baselines.}
\label{fig:result_1}
\end{figure}

\subsubsection{Case Study II: Multimodal Beam Prediction}
We first analyze the dynamic trends of GT beam indices to characterize the dataset's inherent challenges. Fig.~\ref{fig:beamindex_vis} visualizes the spatio-temporal evolution of beam indices under both LoS and NLoS scenarios. 
{\bl As shown in Fig.~\ref{fig:beamindex_vis}(a), the global beam index exhibits periodic sharp jumps despite generally following the environment-affected UAV trajectory. These discontinuities are artifacts of \eqref{eq:global_beamindex}, which maps a continuous 3D spatial tuple into a 1D array, triggering abrupt index shifts at zenith-angle or azimuth boundaries. 
In contrast, NLoS scenarios (Fig.~\ref{fig:beamindex_vis}(b)) manifest highly irregular spatial discontinuities driven by both UAV mobility and environmental scatterers. These observations underscore the necessity of multimodal environmental awareness for wireless sensing and communications.}

\begin{figure}[t]
\centering
\includegraphics[width=0.6\linewidth]{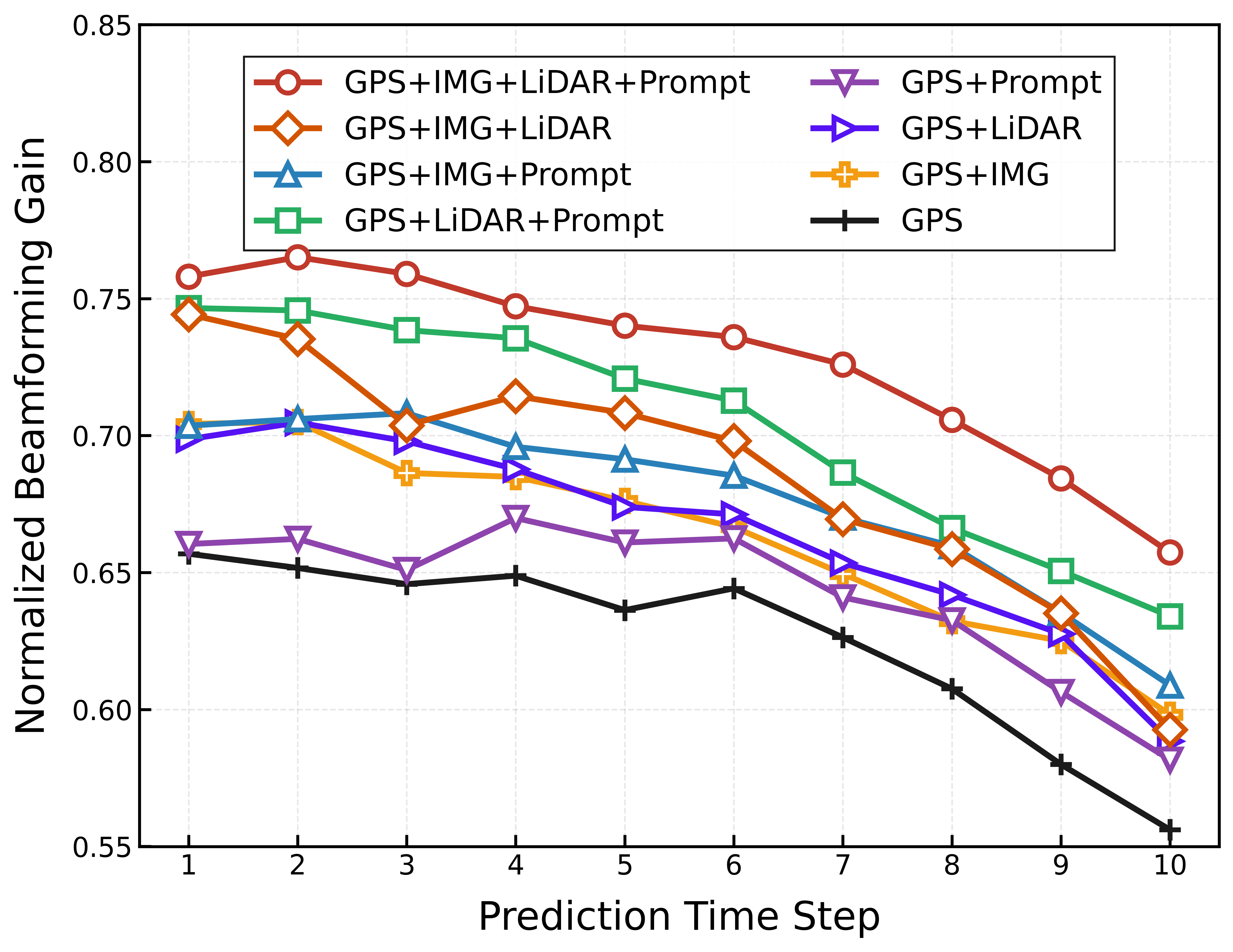}
\caption{{\bl Ablation study on different input modalities.}}
\label{fig:result_2}
\end{figure}

{\bl 
To handle the large search space and avoid the discontinuities of the 1D global beam index, the evaluated LLM-based beam prediction method~\cite{Li2026BeamLLM} predicts the decomposed spatial indices $(k_\theta,k_\varphi,k_r)$ using $T_h=10$ historical multimodal samples and predicts the next $T_p=10$ slots.
We further compare it with far-field beam training, two-stage near-field beam training~\cite{wu2024two}, and the exhaustive-search upper bound.}
As shown in Fig.~\ref{fig:result_1}, the two-stage near-field baseline consistently outperforms the far-field baseline, confirming the importance of near-field modeling for XL-MIMO. Meanwhile, the LLM-based method approaches the exhaustive-search upper bound across all scenarios while avoiding additional online beam-sweeping overhead.
{\bl Finally, the ablation study in Fig.~\ref{fig:result_2} further verifies the contribution of each input modal to beam prediction. GPS-only provides the basic trajectory-aware prediction capability by capturing the historical UAV motion trend. Building on this baseline, adding RGB or LiDAR improves the normalized beamforming gain by introducing visual and geometric environmental cues, while the prompt further contributes trajectory-mode and task-related prior information. The best performance is achieved when GPS, RGB, LiDAR, and prompt are jointly used, demonstrating their complementary roles in environment-aware beam prediction.}

\section{Conclusion}
\label{sec:Conclusion}
In this letter, we presented Multimodal-NF, an open-source wireless-centric multimodal dataset and generation framework for low-altitude near-field XL-MIMO sensing and communications in the upper midband. The proposed dataset provides high-fidelity near-field CSI, synchronized multimodal data, and wireless labels, enabling the evaluation of environment-aware wireless tasks in 3D scenarios. Representative case studies on near-field localization and multimodal beam prediction validate the effectiveness of the proposed dataset and highlight the benefits of multimodal environmental information.

 \small\bibliographystyle{./bibliography/IEEEtran}
 \bibliography{ref}

\vspace{12pt}

\end{document}